\RequirePackage{etoolbox}
\csdef{input@path}{%
 {sty/}
 {img/}
}%
\csgdef{bibdir}{bib/}

\documentclass[ba]{imsart}
\usepackage{amsthm}
\usepackage{amsmath}
\usepackage{natbib}
\usepackage[colorlinks,citecolor=blue,urlcolor=blue,filecolor=blue,backref=page]{hyperref}
\usepackage{graphicx}
\usepackage{rotating}

\startlocaldefs
\endlocaldefs

\begin{document}


\begin{frontmatter}
\title{Bayesian Quantile Regression for Ordinal Longitudinal Data}

\address{ {\small\small } Statistics Department, College of Administration and Economics, Al-Qadisiyah University, Al Diwaniyah, Iraq}
\runtitle{Quantile  Regression}

\begin{aug}
\author{\fnms{Rahim Alhamzawi$^*$} \snm{}}

\runauthor{Rahim Alhamzawi}


\end{aug}

\begin{abstract}
Since the pioneering work by \cite{koenker1978regression}, quantile regression models and its applications have become
increasingly popular and important for research in many areas.
In this paper, a random effects ordinal quantile regression model is proposed for analysis of longitudinal data with ordinal outcome of interest.  An efficient Gibbs sampling algorithm
was derived for fitting the model to the data based on  a location-scale mixture representation of
the   skewed double exponential distribution. The proposed approach is illustrated using simulated data and a real data example. This is the first work to
discuss  quantile regression  for analysis of longitudinal data with ordinal outcome.

\end{abstract}

\begin{keyword}
\kwd{ Bayesian inference, Cut-points, Longitudinal data, Ordinal regression,  Quantile regression }
\end{keyword}

\end{frontmatter}

\section{Introduction}
The mean value has long been used as a measure of location of the center of a distribution.
However, in many applications, there are occasions when  analysts are interested to observe and analyze different points of
the distribution. The distribution function $F$ of a random  variable can be characterized by infinite number of points spanning
its support. These points are called quantiles.
Thus, quantiles are points taken at regular intervals from   $F$. The $\theta$th quantile of a data distribution, $\theta\in(0, 1)$, is  interpreted  as the value such
that there is $100(1-\theta)\%$ of mass on its right side and $100\theta\%$ of mass on its left side. In particular, for a continuous random variable $Y$, the $100\theta\%$ quantile  of $F$ is the value $y$ which solves $F(y)=\theta$ (we assume that this value is unique), where $F(y)=P(Y\leq y)$. Thus, the population lower quartile, median and upper quartile are the solutions to the equations $F(y)=\frac{1}{4}$, $F(y)=\frac{1}{2}$ and $F(y)=\frac{3}{4}$, respectively. Compared to mean value, quantiles are useful measures because they are less susceptible to skewed distributions and outliers. This fact form the building block
of quantile regression which unlike its standard mean regression counterpart, lies in its
flexibility and ability in providing a more complete investigation of the entire conditional distribution
of the response variable distribution given its predictors. To this end, quantile regression has become increasingly popular since the pioneering research  of \cite{koenker1978regression}.

After its introduction, quantile regression has attracted considerable attention in recent literature. It has been applied in a wide range of fields such as agriculture \citep{kostov2013quantile},  body mass index \citep{bottai2014use}, microarray study \citep{wang2007detecting}, financial portfolio \citep{mezali2013quantile},  economics \citep{hendricks1992hierarchical}, ecology \citep{cade2003gentle}, climate change \citep{reich2012spatiotemporal}, survival
analysis \citep{koenker2001reappraising}  and so on.
A comprehensive account of other recent applications of quantile regression can be found in \cite{yu2003quantile} and \cite{koenker2005quantile}.

Longitudinal data is encountered in a wide variety of applications, including economics,  finance,  medicine, psychology and sociology. It is repeatedly measured  from independent subjects  over time and correlation arises between
measures from the same subject.
Since the pioneering work by \cite{laird1982random}, the  mixed models with random
effects   have become  common  and active models to deal with longitudinal data. A number of books and a vast number of research papers published in this area have been motivated by Laird and Ware's mixed models. The majority of these books and  research papers focuses on  standard mean regression. See for example, \cite{wolfinger1993generalized}, \cite{verbeke1996linear}, \cite{hedges1998fixed}, \cite{tao1999estimation}, \cite{mcculloch2001generalized}, \cite{hedeker2006longitudinal} and  \cite{baayen2008mixed}, among others. In contrast, limited   research papers have been
conducted on quantile regression  for longitudinal data. For example,  \cite{koenker2004quantile}  proposed the $l_1$ regularization quantile regression model,
 \cite{lipsitz1997quantile} studied quantile regression for longitudinal data in different contexts and  developed resampling
approaches for inference.
\cite{geraci2007quantile} suggested a Bayesian quantile regression method for longitudinal data using the skewed Laplace distribution  (SLD) for the errors, \cite{reich2010flexible} proposed a flexible Bayesian quantile regression  method  for dependent and independent data using  an infinite
mixture of normals for the errors,  \cite{yuan2010bayesian} studied  quantile regression for longitudinal data with nonignorable intermittent missing data  and \cite{alhamzawi2014bayesian} proposed a method for regularization in mixed quantile regression models.

Longitudinal data with  ordinal responses  routinely appear in many applications,   including economics, psychology and sociology. Existing approaches in classical mean regression are typically designed to deal with such data. At present time, the most common method  in classical mean regression for modelling such data  employs the cumulative logit model. There exists a large literature on the analysis of longitudinal data with  ordinal responses, and we refer to  \cite{fitzmaurice2012applied} for an overview.  In contrast, quantile regression  approaches for estimating the parameters
of ordinal longitudinal data have not been proposed, yet. The goal of this paper is to fill this gap by introducing an ordinal random effects quantile regression model that is appropriate for such data.

In Section \ref{Methods}, we present  a random effects ordinal quantile regression model  for analysis of longitudinal data with ordinal outcome  by using a data augmentation method.  We  also discuss prior elicitation.  In Section \ref{Gibbs Sampler}, we  outline the Bayesian  estimation method via Gibbs sampler.  In Section \ref{Simulation_studies}, we carry
out simulation  scenarios    to investigate  the performance of the proposed method, and in Section \ref{RD}, we illustrate our proposed method using a real dataset.  We conclude the paper with a brief  discussion in Section \ref{conclusion}.  An appendix contains the Gibbs sampler details.

\section{Methods \label{Methods}}
\subsection{Quantile Regression}
Given training data $\{(\mbox{\boldmath$x$}_i, y_i), i=1,\cdots,N\},$ with covariate vector $\mbox{\boldmath$x$}_i\in R^p$ and outcome of interest $y_i\in R$. The $\theta$th quantile regression model  for    the response $y_i$ given the covariate vector $\mbox{\boldmath$x$}_i$ takes the form of
 \begin{equation}\label{EQ1}
 Q_{y_i}(\theta|\mbox{\boldmath$x$}_i)=\mbox{\boldmath$x$}_i'\mbox{\boldmath$\beta$},
 \end{equation}
 where $Q_{y_i}(\theta|\mbox{\boldmath$x$}_i)= F^{-1}_{y_i}(\theta|\mbox{\boldmath$x$}_i)$ is the inverse Cumulative  Distribution Function (CDF) and $\mbox{\boldmath$\beta$}$ is the unknown quantile coefficients vector.
This is what makes  the quantile estimators  can be considered nonparametric maximum likelihood estimators \citep{wasserman2006all}.

 Unlike the standard mean regression,  the error term  does not appear in (\ref{EQ1}) because all the random variation in the conditional  distributions is accounted for by variation
in the $\theta$th quantile, $\theta\in(0, 1)$. Consequently, quantile regression does not require any assumption about the distribution of the errors and, unlike standard mean  regression, is more robust to outliers and non-normal errors, offering  greater statistical efficiency if  the data have outliers or is non-normal.
 It belongs to a robust model family, which can provide a more complete picture of the predictor  effects at different quantile levels of the response variable rather than focusing solely on the  center of the distribution \citep{yu2003quantile}. 
One attractive feature of quantile regression is that  the linear quantile regression model \citep{koenker1978regression} can be used to estimate the parameters of the nonlinear model because  the quantile
regression estimators are equivariant to linear or nonlinear monotonic transformations of the response  variable, i.e., $Q_{\log_{10}y_i}(\theta|\mbox{\boldmath$x$}_i)=\log_{10}Q_{y_i}(\theta|\mbox{\boldmath$x$}_i)$. In general,  this is a very important property since it tells us that  quantile regression provides consistent back transformation and easy  in interpretation in the case of transformations such as the logarithm and the square
root.

 The quantile estimators  have the same interpretation as those of a standard mean regression model except for the indexed  quantile levels where each is estimated \citep{cade2008estimating}. For example,   if the slope is -0.78 for the response variable $y$ given the predictor $x$ in the $95$th quantile would indicate that the $95$th quantile of the response variable  decreased  by 0.78  for each 1 unit increase in $x$. The unknown  quantity  $\mbox{\boldmath$\beta$}$ is estimated by

\begin{eqnarray}\label{EQ2}
\min_{\mbox{\boldmath$\beta$}}\sum_{i=1}^N \rho_{\theta}(y_i-\mbox{\boldmath$x$}_i'\mbox{\boldmath$\beta$}),
\end{eqnarray}
 where $\rho_{\theta} (t)= t\theta - tI(t < 0)$ is the check loss function (CLF) at a quantile $\theta$, $0 < \theta < 1$,
and $I(.)$ is the indicator function. By contrast, standard mean regression method based on the quadratic loss
$t^2$. \cite{koenker1978regression} observed that the CLF (\ref{EQ2}) is closely related
to the skewed Laplace distribution (SLD) and consequently the unknown quantity $\mbox{\boldmath$\beta$}$ can be
estimated through exploiting this link. This observation opens new avenues when dealing with quantile regression and its applications. The density function of a SLD  is
\begin{eqnarray}\label{EQ3}
f(\varepsilon| \theta)={\theta(1-\theta)}\exp\Big\{ -{\rho_\theta(\varepsilon)}\Big\}, \quad \quad \qquad -\infty<\varepsilon<\infty.
\end{eqnarray}
  Minimizing  the CLF (\ref{EQ2}) is equivalent to maximizing
the likelihood function of $\varepsilon_i=y_i-\mbox{\boldmath$x$}_i'\mbox{\boldmath$\beta$}$ by assuming $\varepsilon_i $ from a SLD.
By utilizing the link between the CLF (\ref{EQ2}) and the SLD, \cite{yu2001bayesian} proposed
a Bayesian framework  for quantile regression using the SLD for the error distribution  and show the propriety of
the conditional distribution  of $\mbox{\boldmath$\beta$}$ under an improper prior distribution.
Unfortunately, the joint posterior distribution under this framework does not have a known tractable form and consequently \cite{yu2001bayesian} update the unknown quntity  $\mbox{\boldmath$\beta$}$  from its posterior using the Metropolis-Hastings (M-H) algorithm.  In this context, \cite{kozumi2011gibbs}  proposes  a simple and efficient Gibbs sampling algorithm for updating $\mbox{\boldmath$\beta$}$  by motivating the SLD as a member of the scale mixture of normals.  If $\varepsilon_i \sim \mathrm{N}((1-2\theta) v_i, 2v_i)$,  then the SLD for $\varepsilon_i$ arises when $v_i$ has an exponential distribution with rate parameter $\theta(1-\theta)$. Under this formulation, \cite{yue2011bayesian} presented a Bayesian framework for structured additive quantile regression models, \cite{luo2012bayesian} developed Bayesian
quantile regression for longitudinal data and \cite{alhamzawi2014bayesian} presented Bayesian  Lasso mixed quantile regression.

\subsection{Modeling Ordinal  Longitudinal Data}
Let $\mbox{\boldmath$y$}= \{ y_{ij}\}(i=1, \cdots,N; j=1, \cdots,n_i)$, where $y_{ij}$ denote
the response for the $i$th subject measured at the $j$th time. Then, the  $\theta$th quantile  regression model for ordinal longitudinal data can be formulated  in terms of an ordinal latent  variable $l_{ij}$ as follows:
\begin{eqnarray}\label{EQ4}
 Q_{l_{ij}|\alpha_i}(\theta|\mbox{\boldmath$x$}_{ij}, \alpha_i)=\alpha_i+\mbox{\boldmath$x$}_{ij}'\mbox{\boldmath$\beta$}
\end{eqnarray}
where $Q_{l_{ij}|\alpha_i}(\theta|\mbox{\boldmath$x$}_{ij}, \alpha_i)$  is the inverse CDF
of  the unobserved latent response $l_{ij}$    conditional on a location-shift random effect $\alpha_i$, $\alpha_i\sim$N$(0, \phi)$, and $\mbox{\boldmath$x$}_{ij}$ is a vector of predictors.  The observed ordinal response $y_{ij}$ is assumed to be related to the unobserved response $l_{ij}$ by

$ y_{ij}=\left\{
   \begin{array}{ll}
     1 & \hbox{ if \ \ \ \quad\quad  $\delta_0<l _{ij}\leq\delta_1$;} \\
     c & \hbox{ if \quad \quad $\delta_{c-1}<l _{ij}\leq\delta_c$,}\qquad c=2, \cdots,C-1;\\
     C & \hbox{ if \quad \quad  $\delta_{C-1}<l _{ij}<\delta_C;$}
   \end{array}
 \right.
$
\\{}\\
where $\delta_0, \cdots, \delta_{C}$  are cut-points whose coordinates satisfy
$  -\infty=\delta_0<\delta_1<\cdots<\delta_{C-1}<\delta_{C}=+\infty.$ Here, $\delta_{c-1}$  and $\delta_c$   are respectively defines the lower and upper bounds  of the interval corresponding to observed outcome $c$.

Assuming that the error $\varepsilon_{ij}$ of the unobserved response $l_{ij}$ has a SLD as in (\ref{EQ3}),
we have $\varepsilon_{ij}= (1-2\theta)v_{ij}+\sqrt{2v_{ij}}\epsilon_{ij}$ \citep{kozumi2011gibbs}. Here, the latent variable $v_{ij}$ follows an exponential distribution with rate parameter $\theta(1-\theta)$, and $\epsilon_{ij}$ follows the standard normal distribution.
Then, the CDF for the c category of the observed response $y_{ij}$ is:
\begin{eqnarray}\label{EQ5}
P\big(y_{ij}\leq c|l_{ij}, \delta_c \big)&&=P\big(l_{ij}\leq \delta_c|\mbox{\boldmath$\beta$}, \alpha_i, v_{ij}\big),\nonumber\\&&
=P\big(\alpha_i+\mbox{\boldmath$x$}_{ij}'\mbox{\boldmath$\beta$}+ (1-2\theta)v_{ij}+\sqrt{2v_{ij}}\epsilon_{ij}\leq \delta_c\big),  \nonumber\\&&
=P\Big(\epsilon_{ij}\leq \frac{\delta_c-\alpha_i-\mbox{\boldmath$x$}_{ij}'\mbox{\boldmath$\beta$}-(1-2\theta)v_{ij}}{\sqrt{2v_{ij}}}\Big), \nonumber\\&&
=\Phi\Big(\frac{\delta_c-\alpha_i-\mbox{\boldmath$x$}_{ij}'\mbox{\boldmath$\beta$}-(1-2\theta)v_{ij}}{\sqrt{2v_{ij}}}\Big),
 \nonumber
\end{eqnarray}
 where $\Phi$ is the standard normal CDF.  Using $\Phi$, we can calculate  $P\big(y_{ij}= c|l_{ij},\delta_{c-1}, \delta_c\big)$ as follows:
\begin{eqnarray}\label{EQ6}
P\big(y_{ij}= c|l_{ij},\delta_{c-1}, \delta_c\big)&&=P\big(\delta_{c-1}<l_{ij}\leq\delta_c| \mbox{\boldmath$\beta$}, \alpha_i, v_{ij}\big),\nonumber\\&&
=\Phi\Big(\frac{\delta_c-\alpha_i-\mbox{\boldmath$x$}_{ij}'\mbox{\boldmath$\beta$}-(1-2\theta)v_{ij}}{\sqrt{2v_{ij}}}\Big)-
\Phi\Big(\frac{\delta_{c-1}-\alpha_i-\mbox{\boldmath$x$}_{ij}'\mbox{\boldmath$\beta$}-(1-2\theta)v_{ij}}{\sqrt{2v_{ij}}}\Big).
 \nonumber
\end{eqnarray}



\subsection{Priors}
Prior distribution selection is an essential  step in any Bayesian inference; however, in the Bayesian paradigm it is particularly crucial as issues  can arise when default prior distributions are used without caution \citep{kinney2007fixed}. For the fixed effects \mbox{\boldmath$\beta$}, a typical choice is to assign  a zero mean normal prior distribution  on each $\beta_k, k=1,2, \cdots,p$,  which  leads to the ridge estimator. However, this prior performs poorly if there are big  differences in the size of fixed effects \citep{griffin2010inference}.
 An generalization   of the ridge prior is a Laplace prior, which is equivalent to the
Lasso model \citep{tibshirani1996regression, bae2004gene}. This prior ‏ has received  considerable attention  in the recent literature (for example see, \citet{bae2004gene,park2008bayesian,hans2009bayesian,li2010bayesian,griffin2010inference}).
In this paper, we assign a Laplace prior on each $\beta_k$ takes the form of
\begin{eqnarray}\label{EQ8}
P(\mbox{\boldmath$\beta$}|\lambda)=\prod_{k=1}^p\frac{\lambda}{2}e^{-\lambda|\beta_k|}, \qquad \qquad \lambda\geq 0.
\end{eqnarray}
According to \cite{andrews1974scale}, the prior  (\ref{EQ8}) can be written as
\begin{eqnarray}\label{EQ9}
\prod_{k=1}^p\frac{\lambda}{2}e^{-\lambda|\beta_k|}=\prod_{k=1}^p\int_0^{\infty}\mathrm{N}(\beta_k; 0, s_k)\mathrm{Exp}(s_k; \lambda^2/2)ds_k.
\end{eqnarray}
From (\ref{EQ9}), it can be seen that we assign a zero-mean normal prior distribution  with unknown variance for each $\beta_k$. We specify   an exponential  prior distributions with rate parameter $\lambda^2/2$ for the variances assuming
they are independent.  Then, we put a gamma prior on $\lambda^2$ with shape parameter $a_1$ and rate parameter $a_2$.
 Since $\alpha_i\sim$N$(0, \phi)$, this motivates us to consider an inverse gamma prior on $\phi$ with shape parameter $b_1$ and scale parameter $b_2$.

Following  \cite{montesinos2015genomic} and \cite{sorensen1995bayesian}, we consider
an order statistics from $U(\delta_0, \delta_C)$ distribution, for the $C-1$ unknown cut-points:
\begin{eqnarray}\label{EQ11}
P(\mbox{\boldmath$\delta$})=(C-1)!\Big(\frac{1}{\delta_{\max}-\delta_{\min}}\Big)^{C-1}I(\mbox{\boldmath$\delta$}\in T),
\end{eqnarray}
where $\mbox{\boldmath$\delta$}= (\delta_0, \delta_1, \cdots, \delta_C)$ and
 $T=\{(\delta_{\min},\delta_1, \cdots, \delta_{\max})|\delta_{\min}<\delta_1<\cdots<\delta_{C-1}<\delta_{\max}\}.    $
Because $l_{ij}\sim$N$(\alpha_i+\mbox{\boldmath$x$}_{ij}'\mbox{\boldmath$\beta$}+(1-2\theta)v_{ij}, 2v_{ij})$ and we observe $y_{ij}=c$ if  $\delta_{c-1}<l_{ij}<\delta_c$,
the  posterior distribution of all the parameters and latent variables is given by
\begin{eqnarray}\label{EQ12}
 P(\mbox{\boldmath$\beta$}, \mbox{\boldmath$\alpha$},  \mbox{\boldmath$l$},  \mbox{\boldmath$\delta$}, \mbox{\boldmath$v$}, \mbox{\boldmath$s$}, \lambda^2, \phi|\mbox{\boldmath$y$})& \propto & P(\mbox{\boldmath$y$}|\mbox{\boldmath$l$}, \mbox{\boldmath$\delta$}) P(\mbox{\boldmath$l$}|\mbox{\boldmath$\beta$}, \mbox{\boldmath$\alpha$}, \mbox{\boldmath$v$}) P(\mbox{\boldmath$\delta$})P(\mbox{\boldmath$v$})\nonumber\\
   &\times& P(\mbox{\boldmath$\beta$}|\mbox{\boldmath$s$})P(\mbox{\boldmath$s$}|\lambda^2)P(\lambda^2)P(\mbox{\boldmath$\alpha$}|\phi)
P(\phi),
\end{eqnarray}
where,
 $\mbox{\boldmath$y$}=(y_{11}, \cdots, y_{N_{n_N}})$,
$\mbox{\boldmath$v$}=(v_{11}, \cdots, v_{N_{n_N}})$, $\mbox{\boldmath$l$}=(l_{11}, \cdots, l_{N_{n_N}})$,
$\mbox{\boldmath$\alpha$}=(\alpha_1, \cdots, \alpha_N)$ and $\mbox{\boldmath$s$}=(s_1, \cdots, s_p)$.

The full conditional  distributions for $\mbox{\boldmath$\beta$}, \mbox{\boldmath$\alpha$},  \mbox{\boldmath$l$},  \mbox{\boldmath$\delta$}, \mbox{\boldmath$v$}$,  $\mbox{\boldmath$s$}, \lambda^2$ and $\phi$ are summarized below and
details of all derivations are provided  in Appendix \ref{appendix}.
\section{Gibbs Sampler \label{Gibbs Sampler}}
Using the data augmentation procedure  as in \cite{albert1993bayesian}, a Gibbs sampling method for the
ordinal quantile regression model with longitudinal data is constructed by updating  $\mbox{\boldmath$\beta$}, \mbox{\boldmath$\alpha$},  \mbox{\boldmath$l$},  \mbox{\boldmath$\delta$}, \mbox{\boldmath$v$}$,  $\mbox{\boldmath$s$}$, $\lambda^2$, and $\phi$ from their full conditional distributions. From (\ref{EQ12}), we can construct  a tractable
 algorithm  for efficient posterior computation that works as follows:
\begin{enumerate}
  \item Sample $v_{ij}$ $(i=1, \cdots,N, j=1, \cdots,n_i)$ from the generalized inverse Gaussian distribution GIG$(\nu, \varrho_1, \varrho_2)$,
where $\varrho_1^2={(l_{ij}-\mbox{\boldmath$x$}_{ij}'\mbox{\boldmath$\beta$}-\alpha_i)^2}/{2}$ and $\varrho_2^2=1/2$.
  \item Sample  $\beta_k   (k=1,2, \cdots, p)$  from N$( \mu_{\beta_k}, \sigma_{\beta_k}^2)$ where
\begin{eqnarray*}
  \sigma_{\beta_k}^{2}&=& \Big(\sum_{i=1}^N\sum_{j=1}^{n_i}\frac{x_{ijk}^2}{2v_{ij}}+\frac{1}{s_k}\Big)^{-1},\\
\mathrm{and}\\
 \mu_{\beta_k} &=&   \sigma_{\beta_k}^{2}\sum_{i=1}^N\sum_{j=1}^{n_i}\frac{\Big({l}_{ijk}-\sum_{h=1, h\neq k}^{p}x_{ijh}\beta_h-\alpha_i - (1-2\theta) v_{ij}\Big )x_{ijk}}{2v_{ij}}.
\end{eqnarray*}
\item Sample $s_k, (k=1,2, \cdots, p)$ from GIG$(0.5, \varrho_1, \varrho_2)$, where
$\varrho_1^2=\beta_k^2$ and $\varrho_2^2=\lambda^2$.
\item Sample $\lambda^2$ from Gamma distribution with shape parameter $p+a_1$ and rate parameter $\sum_{k=1}^p s_k/2 +a_2$.
\item Sample $\alpha_i   (i=1, \cdots, N)$ from N$(\mu_{\alpha_i}, \sigma_{\alpha_i}^2)$,  where
\begin{eqnarray*}
 \sigma_{\alpha_i}^{2}&=& \Big(\sum_{j=1}^{n_i}\frac{1}{2v_{ij}}+\frac{1}{\phi}\Big)^{-1},\qquad\qquad\qquad\qquad\qquad\qquad\qquad\\
\mathrm{and}\\
\mu_{\alpha_i} &=&   \sigma_{\alpha_i}^{2}\sum_{j=1}^{n_i}\frac{\Big({l}_{ij}-\mbox{\boldmath$x$}\acute{}_{ij}\mbox{\boldmath$\beta$} - \xi v_{ij}\Big)}{2v_{ij}}.
\end{eqnarray*}
\item Sample $\phi$ from inverse Gamma distribution with shape parameter $ \frac{N}{2}+b_1$ and scale parameter $\sum_{i=1}^{N}\frac{\alpha_i^2}{2}+b_2$.
\item Sample $l_{ij} (i=1, \cdots,N, j=1, \cdots,n_i)$ from truncated normal (TN) distribution  TN$_{(\delta_{c-1},\delta_c)}\Big(\alpha_i+\mbox{\boldmath$x$}_{ij}'\mbox{\boldmath$\beta$}+(1-2\theta)v_{ij}, 2 v_{ij}\Big)$.
\item Sample $\delta_c$ from a uniform distribution on the interval $\big[\min\{\min (l_{ij}|y_{ij}=c+1), \delta_{c+1}, \delta_{\max}\},    \max\{\max (l_{ij}|y_{ij}=c), \delta_{c-1}, \delta_{\min}\}\big]$.
\end{enumerate}
 The details of the proposed Gibbs sampler algorithm and  fully conditional posterior distributions are given in Appendix A.

\section{Simulation Studies \label{Simulation_studies}}
 We carry out a Monte Carlo simulation studies  to assess the performance of the proposed method.  We compared the proposed Bayesian quantile regression method for ordinal longitudinal data, referred to as ``BQOL",  with Bayesian Quantile Regression for Ordinal Models (BQROR) reported by \cite{rahman2016bayesian}.
 The results of Bayesian logistic ordinal regression (BLOR) for longitudinal   data and the maximum likelihood logistic ordinal regression (MLE)  were also reported.  Models were assessed based on the relative average bias  and the  estimated relative efficiency. Suppose that we are interested in the estimation of a vector of
parameters $\mbox{\boldmath$\psi$}\acute{}=(\psi_1, \psi_2, \cdots, \psi_m)$. Then, the relative average bias of $\psi_{h} (h=1, 2, \cdots,m)$ is given by
\begin{eqnarray}
\widehat{\mathrm{bias}}(\hat{\psi}_{h})=\frac{1}{M}\sum_{r=1}^{M}\frac{\hat{\psi}^{r}_{h}-\psi_{h}}{\mid\psi_{h}|}, \nonumber
\end{eqnarray}
and the  estimated relative efficiency
\begin{eqnarray}
\widehat{\mathrm{eff}}_{\mathrm{model}}(\hat{\psi}_{h})=\frac{S^2_{\mathrm{model}}(\hat{\psi}_{h})}{S^2_{\mathrm{BQOL}}(\hat{\psi}_{h})} ,\nonumber
 \end{eqnarray}
where $M$ denotes the number of replications, $\hat{\psi}^{r}_{h}$ is the   parameter  estimate for the $r$th replication, $\psi_{h}$ is the true value,
$S^2(\hat{\psi}_{h})=\frac{1}{M}\sum_{r=1}^{M}(\hat{\psi}^r_{h}-\bar{\psi}_{h})^2$ and $\bar{\psi}_{h}=\frac{1}{M}\sum _{r=1}^{M}\hat{\psi}^r_{h}$.

\subsection{Simulation  1 \label{Simulation_study_1}}
Here, we follow the same simulation strategy introduced by \cite{montesinos2015genomic}. Specifically,
we simulated data from the following liability:
\begin{eqnarray*}
l_{ij}= \beta_1x_{1ij}+\beta_2x_{2ij}+ \beta_3x_{3ij}+\varepsilon_{ij},  \qquad (i=1,\cdots,40; j=1, \cdots, n_i)
\end{eqnarray*}
where $x_{1ij}$ and $x_{2ij}$ were sampled independently from a uniform distribution on the interval $[-0.1, 0.1]$,  $(\beta_1, \beta_2, \beta_3)=(-5,-10,15)$ and $\varepsilon_{ij}$ were sampled independently from a  logistic distribution with location parameter $\mu=0$ and scale parameter $s=1$. Three  values of  $n_i$ were considered, $ n_i= 5,
10, $ and $20$.
The cut-points used were $\delta_1=-0.8416, \delta_2=-0.2533, \delta_3=0.2533$  and $\delta_4=0.8416$. Then the outcome $y_{ij}$ was sampled according to:

$ y_{ij}=\left\{
   \begin{array}{ll}
     1 & \hbox{ if \qquad\ {}  \quad $-\infty<l _{ij}\leq -0.8416$,} \\
     2 & \hbox{ if \quad \quad $-0.8416<l _{ij}\leq -0.2533$,} \\
     3 & \hbox{ if \quad \quad $-0.2533<l _{ij}\leq \ {} \ 0.2533$,} \\
     4 & \hbox{ if \quad \quad\quad $0.2533<l _{ij}\leq \ {} \ 0.8416$,} \\
     5 & \hbox{ if \quad \quad\quad $0.8416<l _{ij}< \ {} \ \infty$.} \\
   \end{array}
 \right.$

For each  choice of $n_i$  $(n_i=5,10,$ and $ 20)$, we generated  200 data sets.
We ran the proposed Gibbs sampler algorithm for 20,000 iterations, after a burn-in period of 2000 iterations.

\begin{table}
\caption{\small {Estimated relative bias and relative efficiency for the simulated data  1. The proposed model (BQOL) is compared with three other models: the Bayesian  quantile regression for ordinal models (BQROR), Bayesian logistic ordinal regression (BLOR) for longitudinal   data and the maximum likelihood logistic ordinal regression (MLE)} \label{Table.Infer1}}
\begin{tabular}{cccccccccccccc}\hline\hline
 & \multicolumn{6}{c}{\qquad \qquad\qquad  BQOL$_{\theta=0.5}$\quad\quad\qquad BQROR$_{\theta=0.5}$} & \multicolumn{5}{c}{\qquad BLOR\quad\qquad\qquad \qquad MLE}   \\
\cline{3-4}
\cline{6-7}
\cline{9-10}
\cline{12-13}
$n_i$             & Parameter & bias & eff &  & bias & eff && bias & eff  &&   bias & eff & \\
\hline
    & $\beta_1$ & -0.051 & 1.000 &&-0.079& 1.103 && 0.082  &1.012&& -0.068 &1.009  \\
    & $\beta_2$ & -0.002 & 1.000 &  & 0.082&1.113&& 0.093  &1.034&&-0.134&  1.071  \\
    & $\beta_3$ & 0.019 & 1.000 &  &  0.169&1.350&& 0.110 &1.131&&0.167& 1.025 \\
5   & $\delta_1$ & -0.012 & 1.000 &  &-0.011 &1.091&& -0.036 &1.113&&-0.031& 1.004  \\
    & $\delta_2$ & 0.003 & 1.000 &  & -0.010&1.014&& -0.009 &1.013&&-0.016& 1.037  \\
    & $\delta_3$ & 0.003 & 1.000 &  & -0.024&1.213&& -0.001 &0.944&&0.002& 1.056  \\
     & $\delta_4$ & 0.012 & 1.000 &  &-0.011 &1.203&& 0.013 &1.147&&0.016&1.007  \\{}\\
    & $\beta_1$ & -0.058 & 1.000 &  & -0.099 & 1.128 && -0.165 & 1.211&& -0.119 & 1.018 \\
    & $\beta_2$ & 0.022 & 1.000 &  &  0.132& 1.059 &  &-0.033&1.067&&-0.100& 1.009 \\
    & $\beta_3$ & 0.020 & 1.000 &  &  0.026 & 1.056 &  &-0.055&1.029&&-0.032&1.035 \\
10   & $\delta_1$ & -0.002 & 1.000 &  &  0.009 & 1.007 &  &-0.001&1.004&&-0.002& 1.021 \\
    & $\delta_2$ & -0.001 & 1.000 &  &  0.004 & 0.999 &  &-0.003&0.998&&0.002&  1.048\\
    & $\delta_3$ & -0.003 & 1.000 &  &  0.038 & 1.006 &  &0.002&1.003&&-0.003& 1.025 \\
     & $\delta_4$ & -0.001 & 1.000 &  &  0.035 & 1.014 & &0.000&0.998&&0.006&  1.008 \\{}\\
    & $\beta_1$ & -0.011 & 1.000 &  & -0.032 & 1.117 && -0.030 & 1.071&& -0.107 & 1.051 \\
    & $\beta_2$ & -0.010 & 1.000 &  &  0.077& 1.019 &  &-0.162&1.131&&-0.116&  1.119\\
    & $\beta_3$ & 0.004 & 1.000 &  &  0.042& 1.107 & &0.029&1.224&&0.108&  1.103 \\
20   & $\delta_1$ & -0.005 & 1.000 &  &  -0.016 & 1.008 & &-0.001&0.999&&-0.003&0.999   \\
    & $\delta_2$ & 0.003& 1.000 &  &  -0.017 & 1.015 &  &0.014&1.007&&-0.004& 1.028 \\
    & $\delta_3$ & 0.000 & 1.000 &  &  -0.029 & 1.073 &  &-0.002&0.996&&-0.016& 1.008 \\
     & $\delta_4$ & 0.003 & 1.000 &  &  -0.036& 1.088 &  &0.002&0.999&&-0.006& 1.034 \\
\hline
\end{tabular}
\end{table}
In Table \ref{Table.Infer1}, we present the simulation results of Simulation study 1 for $\beta_1, \beta_2, \beta_3, \delta_1,$ $ \delta_2, \delta_3$ and $\delta_4$, including the estimated relative bias and the estimated relative efficiency. In general, it  can be seen  that the absolute bias obtained by the  proposed model (BQOL) when $\theta=0.5$ is much smaller  than its competing models. In most cases, BQOL was better than the other methods in terms of bias and the relative efficiency.
The results  suggest that our method performs well compare to other approaches.
 We see that the Bayesian quantile regression approach for ordinal data (BQROR) performs poorly compared to the other methods because it ignores the nature of the  longitudinal data. We also see that as $n_i$ increases, the Bayesian logistic ordinal regression (BLOR) for longitudinal   data yields low bias and more efficiency.
In Table \ref{Table.Infer2},  we present the simulation results of Simulation study 1 for Bayesian quantile regression methods when $\theta=0.25$ and 0.75. Again, in most cases, BQOL was better than BQROR in terms of bias and the relative efficiency.

\newpage
\begin{table}
\caption{\small {Estimated relative bias and relative efficiency  of the Bayesian quantile regression  methods for the simulated data  1 when $\theta=0.25$ and 0.75. } \label{Table.Infer2}}
\begin{tabular}{cccccccccccccc}
\hline\hline
 & \multicolumn{6}{c}{\qquad \qquad\qquad  $\theta=0.25$} & \multicolumn{5}{c}{\qquad $\theta=0.75$}   \\
\cline{3-7}
\cline{9-13}
 & \multicolumn{6}{c}{\qquad \qquad\qquad  BQOL\quad\quad\qquad BQROR} & \multicolumn{5}{c}{\qquad BQOL\quad\qquad\qquad \qquad BQROR}   \\
\cline{3-4}
\cline{6-7}
\cline{9-10}
\cline{12-13}
$n_i$             & Parameter & bias & eff &  & bias & eff && bias & eff  &&   bias & eff & \\
\hline
    & $\beta_1$ &0.128 &1.000&&0.194&1.738&&-0.174&1.000&&-0.191& 1.223 \\
    & $\beta_2$ & -0.057&1.000&&-0.058&2.014&&-0.070&1.000&&0.091& 2.031 \\
    & $\beta_3$ &0.016 &1.000&&0.018&1.374&&0.083&1.000&&0.066& 1.319\\
5   & $\delta_1$ &0.035 &1.000&&-0.030&1.238&&-0.003&1.000&&-0.014&  1.011\\
    & $\delta_2$ &0.018 &1.000&&-0.039&1.098&&0.005&1.000&&-0.059&  1.015 \\
    & $\delta_3$ &0.013 &1.000&&-0.021&1.043&&0.010&1.000&&-0.040&  1.056\\
     & $\delta_4$ &0.020 &1.000&&-0.025&1.051&&0.022&1.000&&-0.027& 1.037\\{}\\
    & $\beta_1$ & -0.132&1.000&&0.137&1.035&&-0.182&1.000&&-0.183& 1.159 \\
    & $\beta_2$ & -0.049&1.000&&0.064&1.836&&-0.054&1.000&&-0.074&  1.037\\
    & $\beta_3$ & -0.003&1.000&&0.002&1.005&&0.011&1.000&&0.083& 1.215\\
10   & $\delta_1$ &-0.016 &1.000&&0.037&1.016&&-0.008&1.000&&0.020& 1.117 \\
    & $\delta_2$ &-0.009 &1.000&&0.040&1.142&&-0.008&1.000&&0.031& 1.027  \\
    & $\delta_3$ &-0.008 &1.000&&0.033&1.115&&-0.007&1.000&&0.004& 1.006 \\
     & $\delta_4$ &0.019 &1.000&&0.036&1.003&&-0.018&1.000&&0.028& 1.014\\{}\\
    & $\beta_1$ & -0.026&1.000&&-0.037&1.731&&0.013&1.000&&0.018& 1.005 \\
    & $\beta_2$ &-0.156 &1.000&&0.210&1.079&&-0.177&1.000&&0.235& 1.371 \\
    & $\beta_3$ &0.037 &1.000&&-0.044&1.117&&0.013&1.000&&0.018& 1.009\\
20   & $\delta_1$ & -0.014&1.000&&0.064&1.008&&-0.012&1.000&&-0.009& 1.004 \\
    & $\delta_2$ &-0.007 &1.000&&0.005&1.116&&-0.009&1.000&&0.016&  1.032 \\
    & $\delta_3$ &0.000 &1.000&&0.030&1.138&&-0.008&1.000&&0.016&  1.098\\
     & $\delta_4$ &0.008 &1.000&&0.070&1.129&&-0.006&1.000&&0.010& 1.007\\
\hline
\end{tabular}
\end{table}

\begin{table}
\caption{\small {The parameter estimations  for the simulated data  2 when $\theta=0.50$ and 0.25. } \label{Table.Infer3}}
\begin{tabular}{cccccccccccccc}
\hline\hline
 & \multicolumn{6}{c}{\qquad \qquad\qquad  BQOL$_{\theta=0.50}$\quad\quad\qquad BQOL$_{\theta=0.25}$} & \multicolumn{5}{c}{ BLOR}   \\
\cline{3-4}
\cline{6-7}
\cline{9-10}
$n_i$             & Parameter & Mean & SD &  & Mean & SD && Mean & SD   \\
\hline
    & $\beta_1$ &-5.169&0.738&&-5.310&1.003&&-5.661&1.069&&& \\
    & $\beta_2$  &-10.028&0.771&&-9.838&1.035&&-9.682&1.036&&&\\
    & $\beta_3$  &15.259&0.688&&15.392&0.935&&14.273&1.053&&&\\
5   & $\delta_1$  &-0.862&0.061&&-0.842&0.083&&-0.871&0.088&&&\\
    & $\delta_2$  &-0.249&0.066&&-0.237&0.064&&-0.231&0.071&&&\\
    & $\delta_3$  &0.251&0.063&&0.261&0.079&&0.261&0.069&&&\\
     & $\delta_4$  &0.849&0.073&&0.857&0.083&&0.881&0.093&&&\\{}\\
    & $\beta_1$ &-5.132&0.935&&-5.116&0.998&&-5.842&1.014&&&\\
    & $\beta_2$  &-10.239&0.722&&-10.117&1.028&&-9.773&0.998&&&\\
    & $\beta_3$  &15.024&0.776&&15.317&0.993&&15.235&1.037&&&\\
10   & $\delta_1$  &-0.848&0.064&&-0.842&0.083&&-0.863&0.087&&&\\
    & $\delta_2$  &-0.258&0.061&&-0.244&0.089&&-0.255&0.073&&&\\
    & $\delta_3$  &0.258&0.073&&0.246&0.076&&0.259&0.079&&&\\
     & $\delta_4$  &0.848&0.059&&0.845&0.037&&0.868&0.081&&&\\{}\\
    & $\beta_1$ &-4.992&0.739&&-5.162&0.893&&-5.337&0.982&&& \\
    & $\beta_2$  &-10.039&0.812&&-10.337&0.926&&-10.241&1.013&&&\\
    & $\beta_3$  &15.040&0.699&&14.893&0.794&&14.753&1.045&&&\\
20   & $\delta_1$  &-0.856&0.066&&-0.853&0.081&&-0.863&0.079&&&\\
    & $\delta_2$  &-0.263&0.085&&-0.257&0.063&&-0.247&0.066&&&\\
    & $\delta_3$  &0.249&0.081&&0.249&0.066&&0.261&0.071&&&\\
     & $\delta_4$  &0.842&0.076&&0.853&0.087&&0.857&0.081&&&\\
\hline
\end{tabular}
\end{table}

\subsection{Simulation  2 \label{Simulation_study_2}}
The setup for this simulation study is the same as  Simulation 1, except we sampled the latent variable $l_{ij}$ as follows:
\begin{eqnarray*}
l_{ij}= \alpha_i+\beta_1x_{1ij}+\beta_2x_{2ij}+ \beta_3x_{3ij}+\varepsilon_{ij},  \qquad \alpha_i\sim \mathrm{N}(0,1).
\end{eqnarray*}
This allows us to examine  the performance of the proposed model   in the case of random effects. In this simulation study, we only consider   the performance of the  Bayesian methods for longitudinal data with ordinal outcome (BQOL and BLOR).  In Table \ref{Table.Infer3}, we present the estimates of the parameters  $\beta_1, \beta_2, \beta_3, \delta_1,$ $ \delta_2, \delta_3$ and $\delta_4$, when $\theta=0.5$ and 0.25. From Table \ref{Table.Infer3} we can see that, our approach  tends to give less biased parameter estimates for  $\beta_1, \beta_2, \beta_3, \delta_1,$ $ \delta_2, \delta_3$ and $\delta_4$ compared to BLOR. The convergence of the proposed Gibbs sampling algorithm in
this simulation study was monitored using the multivariate potential scale reduction factor
(MPSRF) reported by  \cite{brooks1998general}.
\begin{figure}[!h]
\centering
\includegraphics[width=80mm,height= 60mm]{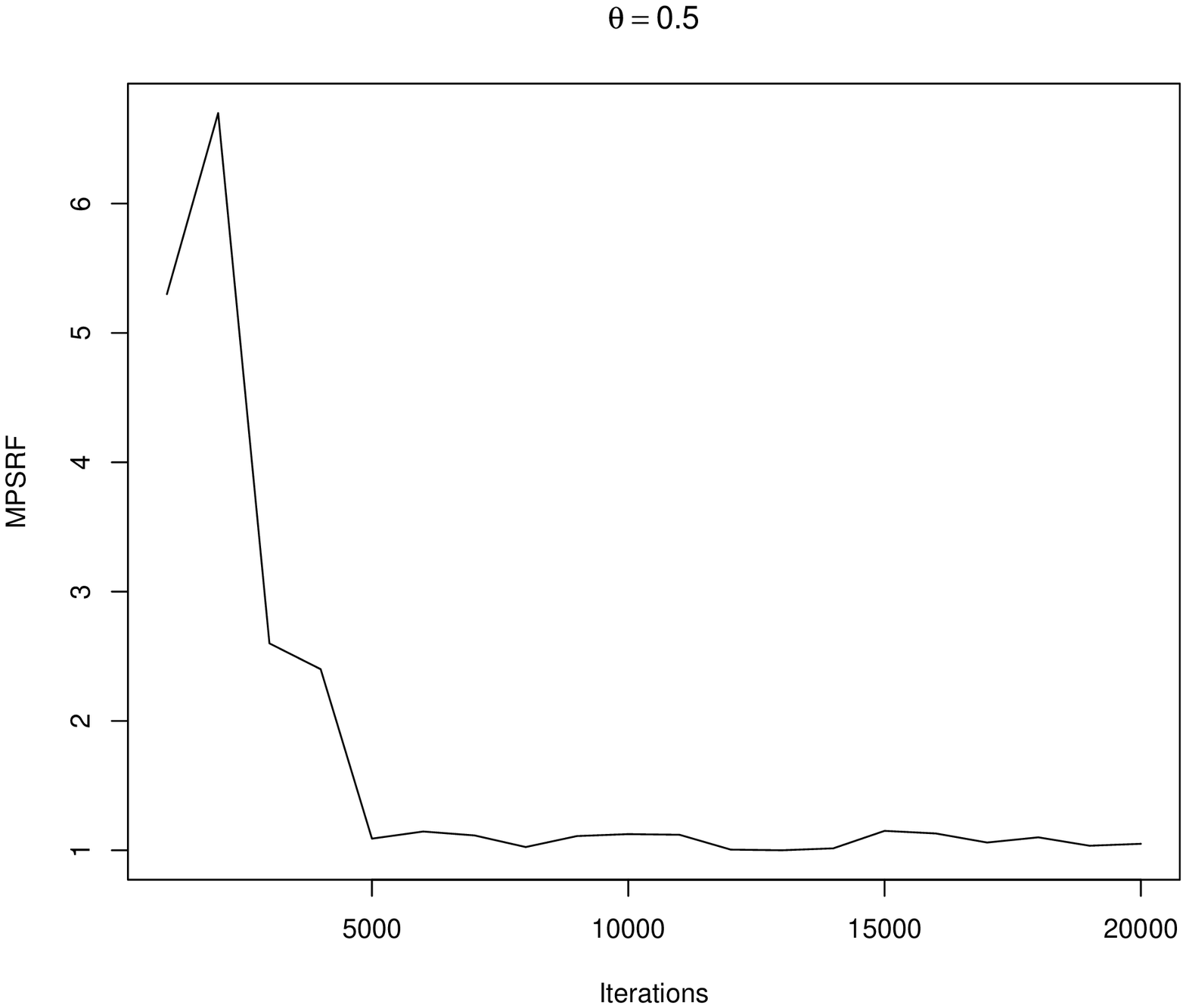}
\includegraphics[width=80mm,height= 60mm]{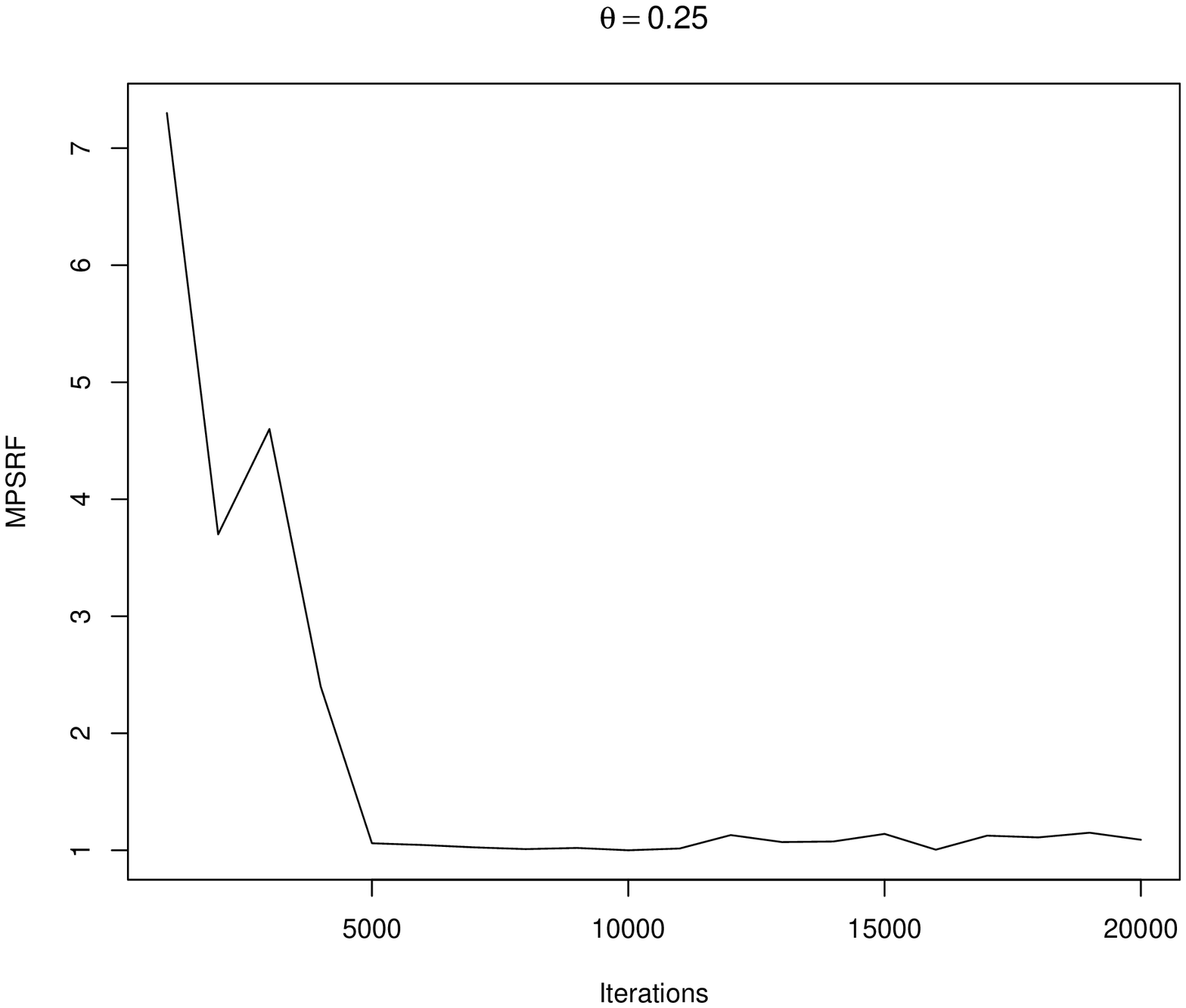}
\caption{MPSRF for longitudinal ordinal quantile regression in Simulation 2.\label{Fig1}}
\end{figure}

From Figure (\ref{Fig1}), it can be observed that the MPSRF becomes stable and very close to 1 after
about the first 5000 iterations for each quantile level under consideration. Hence, the convergence to the posterior distribution was quick and the mixing was good.

\section{Longitudinal Data Example}
In this section, we consider a data set  from  the National Institute of Mental Health Schizophrenia Collaborative  (NIMHSC) study  previously analysed by \citep{gibbons1994application}. The objective of this study is to
 assess treatment-related changes in illness severity over time.
 Specifically, we studied  item 79 (imps79o; severity of illness) of the inpatient
multidimensional psychiatric scale. This item was  measured on a seven point scale as in Table \ref{seven1}:

\begin{center}\label{PointScale}
 \begin{table}
\caption{ Seven point scale for the NIMHSC study      \small    \label{seven1} }
\begin{tabular}{| l | l |}
    \hline
     1 & normal \\ \hline
     2 & borderline mentally ill \\ \hline
    3 & mildly ill \\ \hline
4&moderately ill\\ \hline
5&markedly ill\\ \hline
6&severly ill\\ \hline
7&among the most extremely ill\\ \hline
    \end{tabular}
\end{table}
\end{center}

\cite{hedeker1994random} recorded the seven point scale into four: (1) not ill or borderline, (2) mildly or moderately, (3) markedly ill, (4) severely or most extremely ill. Patients were randomized to receive one of four medications, either placebo or one of three different anti-psychotic drugs. This study consists of three predictors:  TxDrug a dummy coded drug effect  variable (0=Placebo, 1=Drug), the square root of the week  (SqrtWeek), and the interaction between  TxDrug and SqrtWeek (TxSWeek). At the $\theta$th ordinal quantile regression, we considered
\begin{eqnarray*}
  Q_{\mathrm{imps79o}_{ij}}(\theta|\cdot) = \alpha_i+\beta_0+\beta_1 \mathrm{TxDrug}_{ij} + \beta_2 \mathrm{SqrtWeek}_{ij} +\beta_3 \mathrm{TxSWeek}_{ij}
\end{eqnarray*}

\newpage
\begin{table}
\caption{\small {The parameter estimations  for the NIMHSC study when $\theta=0.50$ and 0.25. } \label{RD}}
\begin{tabular}{cccccccccccccc}
\hline\hline
&BQOL$_{\theta=0.50}$&&&BQOL$_{\theta=0.25}$&&&BLOR&&&\\
\cline{2-3}
\cline{5-6}
\cline{8-9}
              Parameter & Mean & $\%95$ CI &  & Mean &  $\%95$ CI && Mean &  $\%95$ CI   \\
\hline
    $\beta_0$ &5.663&(5.037, 6.1192)&&3.527&(2.442, 4.735)&&6.021& (5.571, 7.421)& \\
    $\beta_1$ &-0.073&(-0.523, 0.419)&&-0.048&(-0.661, 0.783)&&-0.227&(-0.991, 0.771)&\\
    $\beta_2$ &-0.746&(-1.351, -0.1337)&&-0.643&(-0.897, -0.437)&&-1.247&(-1.562, -0.661)&\\
    $\beta_3$&-1.206&(-1.663, -0.881)&&-1.104&(-1.437, -0.789)&&-0.883&(-1.641, -0.291)&\\
    $\delta_1$&2.751&(2.538, 2.879)&&2.865&(2.479, 3.125)&&1.997&(1.087, 3.114)&\\
    $\delta_2$&4.173&(3.894, 4.337)&&3.984&(3.716, 4.118)&&3.764&(3.221, 4.261)&\\
    $\delta_3$&5.889&(5.103, 6.709)&&5.765&(5.042, 6.327)&&5.443&(4.793, 6.431)&\\
\hline
\end{tabular}
\end{table}

Table \ref{RD} lists parameter estimations obtained using the Bayesian methods (BQOL and BLOR). The methods are assessed based on $ 95\% $ credible intervals and the deviance information criterion \citep[DIC;][]{spiegelhalter2002bayesian}.
 Clearly, it can be seen that the credible intervals (95\% CI) obtained using the BQOL when $\theta=0.50$ are generally shorter than the credible intervals  obtained using the BLOR, suggesting an efficiency gain and stable estimation
from the posterior distributions. In addition, DIC was computed for our model when $\theta=0.50$ and $\theta=0.25$ as well as for BLOR and the numbers were 3311.32, 3615.48 and 3417.39, respectively. Hence, under $\theta=0.50$, model comparison using DIC indicates  that quantile ordinal models can give a better model fit  compared to the  Bayesian logistic ordinal regression (BLOR) for longitudinal   data. This shows that the model uesd for the errors  in (\ref{EQ3}) is a working model with artiﬁcial assumptions, employed on the outcome variable to achieve the equivalence between
maximising SLD and the minimising proplem in (\ref{EQ2}).

\section{Conclusion \label{conclusion}}
 Since Bayesian quantile methods  for estimating  ordinal models with longitudinal data have not been proposed, yet.
 This paper fills this gap and presents a random effects ordinal quantile regression model  for analysis of longitudinal data with ordinal outcome of interest.  An efficient Gibbs sampling algorithm
was derived for fitting the model to the data based on  a location-scale mixture representation of
the   skewed double exponential distribution. The proposed approach is illustrated using simulated data and a real data example. Results show that the proposed approach performs well. One of the most  desirable features
of the proposed method is its model robustness in the sense that  makes very minimal assumptions on the form of the error term distribution and thus is able to accommodate non-normal errors and outliers, which are popular in many real  world applications.

\appendix
 \section{ Gibbs Sampler Details}\label{appendix}
The full conditional distribution of each $v_{ij}$, denoted by $P(v_{ij}|l_{ij},\mbox{\boldmath$\beta$}, \alpha_i)$ is proportional to $P(l_{ij}|v_{ij},\mbox{\boldmath$\beta$}, \alpha_i)P(v_{ij})$. Thus, we have
\begin{eqnarray*}
P(v_{ij}|l_{ij},\mbox{\boldmath$\beta$}, \alpha_i) &\propto & v_{ij}^{-1/2} \exp\Big\{ -\frac{1}{2}\Big(\frac{l_{ij}-\mbox{\boldmath$x$}_{ij}'\mbox{\boldmath$\beta$}-\alpha_i-\xi v_{ij}}{\sqrt{2v_{ij}}}\Big)^2-\zeta v_{ij}\Big\} \\
   &\propto & v_{ij}^{-1/2} \exp\Big\{ -\frac{1}{2}\Big(\frac{(l_{ij}-\mbox{\boldmath$x$}_{ij}'\mbox{\boldmath$\beta$}-\alpha_i)^2+\xi^2 v_{ij}^2 -2\xi v_{ij}(l_{ij}-\mbox{\boldmath$x$}_{ij}'\mbox{\boldmath$\beta$}-\alpha_i)}{{2v_{ij}}}+2\zeta v_{ij}\Big)\Big\} \\
&\propto & v_{ij}^{-1/2} \exp\Big\{ -\frac{1}{2}\Big[\frac{(l_{ij}-\mbox{\boldmath$x$}_{ij}'\mbox{\boldmath$\beta$}-\alpha_i)^2}{2}v_{ij}^{-1}+\Big(\frac{\xi^2}{2}  +2\zeta \Big) v_{ij}\Big]\Big\} \\
&\propto & v_{ij}^{-1/2} \exp\Big\{ -\frac{1}{2}\Big[\frac{(l_{ij}-\mbox{\boldmath$x$}_{ij}'\mbox{\boldmath$\beta$}-\alpha_i)^2}{2}v_{ij}^{-1}+\Big(\frac{1}{2}  \Big) v_{ij}\Big]\Big\} \\
&\propto & v_{ij}^{-1/2} \exp\Big\{ -\frac{1}{2}\Big[\varrho_1^2 v_{ij}^{-1}+\varrho_2^2   v_{ij}\Big]\Big\}, \\
\end{eqnarray*}
where $\xi=1-2\theta$ and $\zeta=\theta(1-\theta)$. Thus, the full conditional distribution of each $v_{ij}$ is a generalized inverse Gaussian distribution GIG $(\nu, \varrho_1, \varrho_2)$,
where $\varrho_1^2={(l_{ij}-\mbox{\boldmath$x$}_{ij}'\mbox{\boldmath$\beta$}-\alpha_i)^2}/{2}$ and $\varrho_2^2=1/2$.
Recall that if $x\sim$ GIG $(0.5, \varrho_1, \varrho_2)$ then the pdf of $x$ is given by \citep{barndorff2001non}
\begin{eqnarray*}
f(x|\nu,\varrho_1,\varrho_2)=\frac{(\varrho_2/\varrho_1)^\nu}{2K_\nu(\varrho_1\varrho_2)}x^{\nu-1}\exp\left\{-\frac{1}{2}
(x^{-1}\varrho_1^2+x\varrho_2^2)\right\},
\end{eqnarray*}
where $x>0, \ -\infty<\nu<\infty, \ \varrho_1,\varrho_2\geq 0$ and $K_\nu(.)$ is so called  ``\emph{modified Bessel function of the third kind}".

The full conditional distribution of each $\beta_k$, denoted by $P(\beta_k|\mbox{\boldmath$l$}, \mbox{\boldmath$\beta$}_{-k}, \mbox{\boldmath$\alpha$}, \mbox{\boldmath$v$},  s_k)$ is proportional to $P(\mbox{\boldmath$l$}|\mbox{\boldmath$\beta$}, \mbox{\boldmath$\alpha$}, \mbox{\boldmath$v$}, s_k)P(\beta_k|s_k)$, where $\mbox{\boldmath$\beta$}_{-k}$ is the vector $\mbox{\boldmath$\beta$}$ excluding the element $\beta_k$. Thus, we have
\begin{eqnarray*}
  P(\beta_k|\mbox{\boldmath$l$}, \mbox{\boldmath$\beta$}_{-k}, \mbox{\boldmath$\alpha$}, \mbox{\boldmath$v$},  s_k)
 &\propto& P(\mbox{\boldmath$l$}|\mbox{\boldmath$\beta$}, \mbox{\boldmath$\alpha$}, \mbox{\boldmath$v$}, s_k)P(\beta_k|s_k) \\
   &\propto& \exp\Big\{-\frac{1}{2}\sum_{i=1}^N\sum_{j=1}^{n_i}\frac{(l_{ij}-\mbox{\boldmath$x$}_{ij}'\mbox{\boldmath$\beta$}-\alpha_i-\xi v_{ij})^2}{2v_{ij}}\Big\} \exp\Big\{-\frac{\beta_k^2}{2s_k}\Big\}\\
   &\propto& \exp\Big\{-\frac{1}{2}\Big[ \Big(\sum_{i=1}^N\sum_{j=1}^{n_i}\frac{x_{ijk}^2}{2v_{ij}}+\frac{1}{s_k})\beta_k^2-2\sum_{i=1}^N\sum_{j=1}^{n_i}\frac{\tilde{l}_{ijk}x_{ijk}}{2v_{ij}}\beta_k\Big] \Big\}, \\
\end{eqnarray*}
where $\mbox{\boldmath$x$}_{ij}=(x_{ij1}, \cdots, x_{ijp})$ and  $\tilde{l}_{ijk}={l}_{ijk}-\sum_{h=1, h\neq k}^{p}x_{ijh}\beta_h-\alpha_i - \xi v_{ij}$. Then the full conditional distribution for $\beta_k$ is normal with mean $\mu_{\beta_k} $ and variance $\sigma_{\beta_k}^{2}$, where
\begin{eqnarray*}
  \sigma_{\beta_k}^{-2}&=& \sum_{i=1}^N\sum_{j=1}^{n_i}\frac{x_{ijk}^2}{2v_{ij}}+\frac{1}{s_k},\\
\mathrm{and}\\
 \mu_{\beta_k} &=&   \sigma_k^{2}\sum_{i=1}^N\sum_{j=1}^{n_i}\frac{\tilde{l}_{ijk}x_{ijk}}{2v_{ij}}.
\end{eqnarray*}
The full conditional distribution of each $s_k$, denoted by $P(s_k|\beta_k)$ is
\begin{eqnarray*}
  P(s_k|\beta_k) &\propto&   P(\beta_k|s_k)P(s_k)\\
   &\propto&  \frac{1}{\sqrt{2\pi s_k}}\exp\Big\{-\frac{\beta_k^2}{2s_k}\Big\}\exp\Big\{ -\frac{\lambda^2}{2} s_k \Big\}\\
&\propto&  s_k^{-\frac{1}{2}}\exp\Big\{-\frac{1}{2}\Big({\beta_k^2}s_k^{-1} +{\lambda^2} s_k \Big)\Big\}\\
\end{eqnarray*}
Thus, the full conditional distribution of $s_k$ is a  GIG$(0.5, \varrho_1, \varrho_2)$, where
$\varrho_1^2=\beta_k^2$ and $\varrho_2^2=\lambda^2$.

The full conditional distribution of  $\lambda^2$, denoted by $P(\lambda^2|\mbox{\boldmath$s$})$ is
\begin{eqnarray}
  P(\lambda^2|\mbox{\boldmath$s$}) &=& P(\mbox{\boldmath$s$}|\lambda^2)P(\lambda^2) \\
   &\propto& \prod_{k=1}^{p}\frac{\lambda^2}{2}\exp\Big\{   -\frac{\lambda^2}{2}s_k\Big\} {(\lambda^2)}^{a_1-1}\exp\{-a_2\lambda^2\}\\
&\propto& (\lambda^2)^{p+a_1-1}\exp\Big\{   -\lambda^2\Big(\sum_{k=1}^p s_k/2 +a_2   \Big)  \Big\}
\end{eqnarray}
That is, the full conditional distribution of $\lambda^2$ is a Gamma distribution.

The full conditional distribution of each $\alpha_i$, denoted by $P(\alpha_i|\mbox{\boldmath$l$}, \mbox{\boldmath$\beta$}, \mbox{\boldmath$v$}, \phi)$ is proportional to $P(\mbox{\boldmath$l$}|\alpha_i, \mbox{\boldmath$\beta$}, \mbox{\boldmath$v$})P(\alpha_i|\phi)$. Thus, we have
\begin{eqnarray*}
  P(\alpha_i|\mbox{\boldmath$l$}, \mbox{\boldmath$\beta$}, \mbox{\boldmath$v$}, \phi)
 &\propto& P(\mbox{\boldmath$l$}|\alpha_i, \mbox{\boldmath$\beta$}, \mbox{\boldmath$v$})P(\alpha_i|\phi)\\
   &\propto& \exp\Big\{-\frac{1}{2}\sum_{j=1}^{n_i}\frac{(l_{ij}-\mbox{\boldmath$x$}_{ij}'\mbox{\boldmath$\beta$}-\alpha_i-\xi v_{ij})^2}{2v_{ij}}\Big\} \exp\Big\{-\frac{\alpha_i^2}{2\phi}\Big\}\\
   &\propto& \exp\Big\{-\frac{1}{2}\Big[ \Big(\sum_{j=1}^{n_i}\frac{1}{2v_{ij}}+\frac{1}{\phi})\alpha_i^2-2\sum_{j=1}^{n_i}\frac{\eta_{ij}}{2v_{ij}}\alpha_i\Big] \Big\}, \\
\end{eqnarray*}
where   $\eta_{ij}={l}_{ij}-\mbox{\boldmath$x$}\acute{}_{ij}\mbox{\boldmath$\beta$} - \xi v_{ij}$. Then the full conditional distribution for $\alpha_i$ is normal with mean $\mu_{\alpha_i} $ and variance $\sigma_{\alpha_i}^{2}$, where
\begin{eqnarray*}
 \sigma_{\alpha_i}^{-2}&=& \sum_{j=1}^{n_i}\frac{1}{2v_{ij}}+\frac{1}{\phi},\\
\mathrm{and}\\
\mu_{\alpha_i} &=&   \sigma_{\alpha_i}^{2}\sum_{j=1}^{n_i}\frac{\eta_{ij}}{2v_{ij}}.
\end{eqnarray*}
The full conditional distribution of  $\phi$, denoted by $P(\phi|\mbox{\boldmath$\alpha$})$, is proportional to $P(\mbox{\boldmath$\alpha$}|\phi) P(\phi)$. Thus, we have
\begin{eqnarray*}
  P(\phi|\mbox{\boldmath$\alpha$}) &\propto&  P(\mbox{\boldmath$\alpha$}|\phi) P(\phi),\\
   &\propto&\Big(\prod_{i=1}^{N}\frac{1}{\sqrt{2 \pi \phi}}\Big) \exp\Big\{-\frac{\sum_{i=1}^{N}\alpha_i^2}{2\phi}\Big\}\phi^{-b_1-1}\exp\Big\{ -\frac{b_2}{ \phi} \Big\}\\
 &\propto&\phi^{-\frac{N}{2}-b_1-1}
\exp\Big\{-\frac{1}{\phi}\Big(\frac{\sum_{i=1}^{N}\alpha_i^2}{2}+b_2 \Big) \Big\}\\
\end{eqnarray*}
That is, the full conditional distribution of $\phi$ is a inverse Gamma distribution.

The full conditional distribution of each $l_{ij}$, denoted by $P(l_{ij}|\mbox{\boldmath$\beta$},\mbox{\boldmath$\delta$}, \alpha_i, v_{ij})$ is proportional to  $P(y_{ij}|l_{ij},\mbox{\boldmath$\delta$})$ $P(l_{ij}|\mbox{\boldmath$\beta$}, \alpha_i, v_{ij} )$.  Thus, we have
\begin{eqnarray*}
 P(l_{ij}|\mbox{\boldmath$\beta$},\mbox{\boldmath$\delta$}, \alpha_i, v_{ij})  &\propto& P(y_{ij}|l_{ij},\mbox{\boldmath$\delta$})P(l_{ij}|\mbox{\boldmath$\beta$}, \alpha_i, v_{ij} ) \\
   &\propto& \textbf{1}\{\delta_{c-1} <l_{ij}\leq\delta_{c}\} \mathrm{N}(l_{ij}; \mbox{\boldmath$x$}_{ij}'\mbox{\boldmath$\beta$}+\alpha_i+\xi v_{ij}, 2v_{ij})
\end{eqnarray*}
That is, the full conditional distribution of $l_{ij}$ is a truncated normal distribution.

At last, the full conditional posterior distribution of $\delta_c$,  denoted by $P(\delta_c|\mbox{\boldmath$y$}, \mbox{\boldmath$l$})$ is proportional to $p(\mbox{\boldmath$y$}|\mbox{\boldmath$l$}, \mbox{\boldmath$\delta$})   P(\mbox{\boldmath$\delta$})$. Thus, we have
\begin{eqnarray}
  P(\delta_c|\mbox{\boldmath$y$}, \mbox{\boldmath$l$}) &\propto& p(\mbox{\boldmath$y$}|\mbox{\boldmath$l$}, \mbox{\boldmath$\delta$})   P(\mbox{\boldmath$\delta$})\\
  &\propto&\prod_{i=1}^N\prod_{j=1}^{n_i}\sum_{c=1}^C \textbf{1}(y_{ij}=c)\textbf{1}(\delta_{c-1}<l_{ij}<\delta_c)\textbf{1}(\mbox{\boldmath$\delta$}\in T)
\end{eqnarray}
Following  \cite{montesinos2015genomic} and \cite{sorensen1995bayesian}, the full conditional distribution of $\delta_c$ is
\begin{eqnarray*}
P(\delta_c|\mbox{\boldmath$y$}, \mbox{\boldmath$l$})=\frac{1}{\min\Big(l_{ij}|y_{ij}=c+1\Big)-\max\Big(l_{ij}|y_{ij}=c\Big)}\textbf{1}(\mbox{\boldmath$\delta$}\in T)
\end{eqnarray*}
That is, the full conditional distribution of $\delta_c$ is a uniform distribution.



\end{document}